\pdfoutput=1
\documentclass[10pt,twocolumn]{IEEEtran}
\usepackage{amsmath,amsthm,amsfonts,amssymb,algorithm,graphicx}

\newtheorem{Theorem}{Theorem}
\newtheorem{Lemma}{Lemma}

\theoremstyle{definition}

\newcommand{\norm}[1]{\| {#1} \| }

\newcommand{\R}{\mathbb{R}}

\newcommand{\tr}{\mathrm{tr}}

\newcommand{\Or}{\mathcal{O}}

\renewcommand{\>}{\right\rangle}

\title{On Sparse Representations of Linear Operators and the Approximation of Matrix Products}
\author{Mohamed-Ali Belabbas and Patrick J. Wolfe\thanks{%
School of Engineering and Applied Sciences,
Department of Statistics, Harvard University,
Oxford Street, Cambridge, MA 02138 USA.  E-mail: \{belabbas,
patrick\}@seas.harvard.edu}}

\begin{document}

\maketitle

\begin{abstract}
Thus far, sparse representations have been exploited largely in the
context of robustly estimating functions in a noisy environment from a
few measurements. In this context, the existence of a basis in which
the signal class under consideration is sparse is used to decrease the
number of necessary measurements while controlling the approximation
error. In this paper, we instead focus on applications in numerical
analysis, by way of sparse representations of linear operators with
the objective of minimizing the number of operations needed to perform
basic operations (here, multiplication) on these operators. We
represent a linear operator by a sum of rank-one operators, and show
how a sparse representation that guarantees a low approximation error
for the product can be obtained from analyzing an induced quadratic
form.  This construction in turn yields new algorithms
for computing approximate matrix products.
\end{abstract}

\section{Introduction}
\label{sec:intro}

\PARstart{O}{perations} on large matrices are a cornerstone of computational linear
algebra.  With a few exceptions such as the approximate Lanczos and
power methods, most algorithms used by practitioners aimed to optimize
speed of computation under the constraint of obtaining an exact
result. Recently, spurred by the seminal paper of Frieze et
al.~\cite{frieze98fast}, there has been a greater interest in finding
algorithms which sacrifice the precision of the result for a gain in
the speed of execution.

Consider the low-rank approximation problem; i.e., finding a matrix
$A_k$ of rank at most $k$ which approximates a given matrix $A$. The
best matrix $A_k$, best in the sense that it minimizes $\norm{A-A_k}$
for any unitarily invariant norm (e.g., spectral or Frobenius norms),
can be obtained by computing the singular value decomposition (SVD) of
$A$. (Throughout this paper, we adopt the Frobenius norm; we use the notation $A_k$ to denote the
best rank-$k$ approximation to $A$ and $\widetilde{A}_k$ to denote an
approximation to it---it will be easy to avoid confusion with $A_i$,
which is used to denote the $i^{\text{th}}$ column of $A$) But in some
instances, evaluating the SVD, which scales as $\Or(n^3)$ where $n$ is
the largest dimension of $A$, may be too costly. Frieze et
al.~in~\cite{frieze98fast} showed that $A_k$ can be reasonably well
approximated by computing the SVD of a subset of the columns of $A$
only, where the columns are sampled according to their relative
powers---i.e., $\operatorname{Pr}(\mbox{pick column $i$}) \varpropto
\norm{A_i}^2/\sum \norm{A_i}^2$, with the expected error coming from using
the approximation $\widetilde{A}_k$ instead of $A_k$ being of the form
$\operatorname{\mathbb{E}}\norm{A-\widetilde{A}_k}^2 \leq
\norm{A-A_k}^2+\varepsilon\norm{A}^2$. In subsequent papers it
has been argued that the additive error term in $\norm{A}^2$ may be
large, and thus other sampling techniques have been introduced to
obtain relative approximation error bounds (see,
e.g.,~\cite{deshpanderandom,sarlos:iaa}).

In this paper, we address the sparse representation of linear operators for the approximation of matrix products. An important object that will appear in our study is the so-called Nystr\"om method (see Section~\ref{sec:background}) to find a low-rank approximation to a positive kernel.  This method, familiar to numerical analysts, has nowadays found applications beyond its original field, most notably in machine learning. In previous work~\cite{belabbas07:_spect_method_machin_learn}, we proposed an approach for low-rank approximation in such applications; here we show that the task of finding optimal sparse representation of linear operators, in order to evaluate their product, is related to the Nystr\"om extension of a certain positive definite kernel. We will the use this connection to derive and bound the error of two new algorithms, which our simulations indicate perform well in practice.

Related to our work is that of~\cite{drineas06fast} for matrix products.  In that paper, Drineas et al.~showed that a randomized algorithm sampling columns $A_i$ and rows $B^j$ of $A$ and $B$ in proportion to their relative powers $\norm{A_i}^2$ and $\norm{B^j}^2$ yields an expected error of
$\operatorname{\mathbb{E}}\norm{AB-\widetilde{AB}}^2 \leq \text{const}\cdot k^{-1} \norm{A}^2\norm{B}^2$. Notice that this bound does not involve a low-rank approximation of $A$ or $B$.  In contrast, we obtain a randomized algorithm bound in which the approximating rank $k$ of a kernel related to $A$ and $B$ appears explicitly.

The methods mentioned above are all \emph{adaptive}, in the sense that they require some knowledge about $A$ and $B$. A very simple non-adaptive method is given by an application of the Johnson-Lindenstrauss Lemma: it is easy to show~\cite{arriaga99algorithmic} that if $W$ is a $k\times n$ matrix with independent unit Normal entries and $x,y \in \R^n$, then for $0 < \epsilon <1$ we have that
\begin{equation*}
\operatorname{Pr}\left(\left| \langle x,y\rangle - k^{-1}\langle Wx,Wy \rangle \right| \leq \epsilon \norm{x}\norm{y} \right)
\geq 1-4e^{-\frac{k}{2}(\frac{\epsilon^2}{2}-\frac{\epsilon^3}{3})}
\text{.}
\end{equation*}
Letting $A^i$ denote the $i^{\text{th}}$ row of $A$ and $B_j$ the $j^{\text{th}}$ column of $B$, we observe the following element-wise relation:
\begin{equation*}
(AB)_{ij} - k^{-1} (AW^TWB)_{ij} = \langle A^i,B_j \rangle - k^{-1} \langle WA^i,WB_j \rangle \text{,}
\end{equation*}
and thus we see that approximating $AB$ by $k^{-1}AW^TWB$ yields a good result with high probability. Later we compare this method to our algorithms described below.

The remainder of this paper is organized as follows. In Section~\ref{sec:background},
we briefly review the Nystr\"om method used to approximate positive
definite matrices~\cite{seeger,fowlkes04spectral}. In
Section~\ref{sec:mat_approx}, we introduce the problem of
approximating a matrix product and highlight two key aspects: the issue of best subset selection and the issue of optimal
rescaling.  We then solve the optimal rescaling problem and analyze a
 randomized and a deterministic algorithm for subset selection; we conclude with simulations and a brief discussion of algorithmic complexity.

\section{Approximation Via The Nystr\"om Method}
\label{sec:background}

To provide context for our results, we first
introduce the so-called Nystr\"om method to approximate the
eigenvectors of a symmetric positive semi-definite (SPSD) matrix.

\subsection{The Nystr\"om Method for Kernel Approximation}

The Nystr\"om method, familiar in the context of finite element
methods, has found many applications in machine learning and computer
vision in recent years (see, e.g.,~\cite{fowlkes04spectral} and
references therein).  We give here a brief overview : Let $k:[0,1]
\times [0,1] \rightarrow \R$ be a positive semi-definite kernel and
$(\lambda_i,f_i)$, $i=0,1,\dots,M$, denote pairs of eigenvalues and
eigenvectors such that
\begin{equation}
  \label{eq:eigkernel}\int_{[0,1]}k(x,y)f_i(y)dy=\lambda_if_i(x).
\end{equation}
The Nystr\"om extension is a method to approximate the eigenvectors of
$k(x,y)$ based on a discretization of the interval $[0,1]$.  Define
the $M+1$ points $x_i$ by $x_m=x_{m-1}+1/M$ with $x_0=0$, so that the
$x_i$'s are evenly spaced along the interval $[0,1]$. Then form the
\emph{Gram matrix} $K_{mn}:=k(x_m,x_n)$, which in turn is used to
approximate~\eqref{eq:eigkernel} by a finite-dimensional spectral
problem
\begin{equation*}
\frac{1}{M+1}\sum_n K_{mn} v_i(n)=\lambda_i^vv_i(m), \quad i=0,1,\ldots,M.
\end{equation*}
The Nystr\"om extension then uses these $v_i$ to give an estimate
$\hat f_i$ of the $i^{\text{th}}$ eigenfunction as follows: $$\hat
f_i(x)=\frac{1}{(M+1)\lambda_i^v}\sum_m k(x,x_m)v_i(m).$$

This method can also be applied in the context of matrices.  Let
$Q$ be an $n \times n$ SPSD matrix, partitioned
as
\begin{equation*}
    Q=\left[\begin{array}{cc} Q_J & Y \\Y^T &
        Z\end{array}\right],
  \end{equation*}
where $Q_J \in \R^{k \times k}$ and $k$ is typically much smaller than
$n$.  It is then possible to approximate $k$ eigenvectors and
eigenvalues of $Q$ by using the eigendecomposition of $Q_J$ as follows.
Define $Q=U\Lambda U^T$ and $Q_J=U_J\Lambda_J{U_J}^T$ with $U,U_J$
orthogonal and $\Lambda,\Lambda_J$ diagonal. The Nystr\"om extension
then tells us that an approximation for $k$ eigenvectors in $U$ is
given by
$$
\widetilde U=\left[\begin{array}{c}U_J
    \\ Y^T U_J{\Lambda_J}^{-1}\end{array}\right].
$$

These approximations $\widetilde{U}\approxeq U$ and $\Lambda_J \approxeq
\Lambda$ in turn yield an approximation $\widetilde Q$ to $Q$ as
follows:
\begin{equation*}
\widetilde Q = \widetilde U \Lambda_J \widetilde U^T=\left[
  \begin{array}{cc} Q_J & Y \\Y^T & Y^T{Q_J}^{-1}Y \end{array}\right].
\end{equation*}
The quality of this approximation can then be measured as the (e.g.,
Frobenius) norm of the Schur complement of $Q_J$ in $Q$:
\begin{equation*}
  \norm{Q-\widetilde Q}=\norm{Z-Y^T{Q_J}^{-1}Y}.
\end{equation*}

\section{Approximation of Matrix Products}
\label{sec:mat_approx}

Let $A\in \R^{m \times n}$ and $B\in\R^{n\times p}$.  We use the
notation $A_i$ to denote the columns of $A$ and $B^i$ the rows of $B$.
We can write the product $AB$ as the sum of rank-one matrices as
follows:
\begin{equation}
  \label{eq:sumrankone} AB=\sum_{i=1}^n A_iB^i.
\end{equation}

Our approach to estimate the product $AB$, akin to model selection in
statistics, will consist
of keeping only a few terms in the sum of~\eqref{eq:sumrankone}; this
entails choosing a subset of columns of $A$ and of rows of $B$, and
rescaling their outer products as appropriate.
To gain some basic insight into this problem, we may consider the following two extreme
cases with $A\in \mathbb{R}^{n \times 2}$ and $B=A^T$.  First, suppose
that the vectors $A_1$ and $A_2$ are collinear.  Then
$AB=A_1A_1^T+\alpha A_1A_1^T=(1+\alpha)A_1A_1^T$, hence we can recover
the product without error by only keeping one term of the sum
of~\eqref{eq:sumrankone} and rescaling it appropriately
\emph{provided} that we know the correlation between $A_1$ and $A_2$.
At the other extreme, if $A_1$ and $A_2$ are orthogonal,
rescaling will not decrease the error no matter which term
in~\eqref{eq:sumrankone} is kept.

Hence we see that there are two key aspects to
the problem of sparse matrix product approximation as formulated above:
\begin{description}
\item[Optimal Model Selection]~\\ Which rows/columns should
  be retained?
\item[Optimal Reweighting]~\\ How should these rows/columns
  be rescaled?
\end{description}

As we show below, the latter of these problems can be solved exactly
for a relatively low complexity.  For the former, which is combinatorial in
nature and seemingly much harder to solve, we give an efficient
approximation procedure.

\section{Solving the Optimal Reweighting Problem}

We first consider the problem of optimal reweighting, conditioned upon
a choice of subset.  In particular, suppose that an
oracle gives us the best subset $J \subset \lbrace 1,\ldots,n\rbrace$
of cardinality $k$ to estimate the product $AB$. Without loss of
generality, we assume that $J=\lbrace 1,\ldots,k\rbrace$. We then have
the following result characterizing how well one can estimate the
product $AB$:

\begin{Theorem}\label{th:maintheo}Let the $n \times n$
  SPSD matrix $Q$ be defined as $Q_{ij} = \langle
  A_i,A_j\rangle\langle B^i,B^j\rangle $ (i.e., $Q=(A^TA) \odot (BB^T)$,
  where $\odot$ is the Hadamard or entrywise product of matrices) and
  have the partition
  \begin{equation}
    \label{eq:partQ}
    Q=\left[\begin{array}{cc} Q_J & Y \\Y^T &
        Z\end{array}\right],
  \end{equation}
  where $J = \lbrace 1,\ldots,k\rbrace$ without loss of generality,
  and $Q_J$ is the corresponding principal submatrix.

  Then the best approximation to the product $AB$ using the terms
  $\{A_iB^i\}_{i \in J}$ is given by
  \begin{equation}
    \label{eq:tildeab}AB \approxeq
    \widetilde{AB}:=\sum_{i\in J}w_iA_iB^i,
  \end{equation}
  where 
 \begin{equation*}
    w:={Q_J}^{-1}r
  \end{equation*}
  and
  \begin{equation*}
  r_i:= \sum_{j=1}^n \langle A_i,A_j\rangle\langle B^i,B^j\rangle,
  \quad i \in J.
  \end{equation*}
  Moreover, if $E$ is the $(n-k) \times (n-k)$ matrix with all entries
  equal to one, then the squared approximation error in Frobenius norm is
  given by
  \begin{equation*}
    \norm{AB-\widetilde{AB}}^2=\tr(S_C(Q_J)E),
  \end{equation*}
  with $S_C(Q_J) := Z-Y^T{Q_J}^{-1}Y$ the Schur complement of $Q_J$ in
  $Q$.
\end{Theorem}

This result tells us how well we can approximate the product $AB$
\emph{granted} that we know only a few rows/columns of $A$ and $B$,
and their correlations with the remaining rows and columns. It also
allows us to characterize the best subset $J$ of size
$k$; it is the subset that minimizes
$\tr(S_C(Q_J)E)$.  

\begin{IEEEproof}
  Given the subset $J$ of $\left\lbrace 1,\ldots,n \right\rbrace$, we
  seek the best scaling factors $w_i$ to minimize the
  squared approximation error $\norm{AB-\widetilde{AB}}^2$.  We can write the
  squared error as
  \begin{multline*} \norm{AB-\sum_{i=1}^k w_i A_i B^i}^2=\\
    \tr\left((AB-\sum_{i=1}^k w_i A_i B^i)^T (AB-\sum_{i=1}^k w_i A_i
      B^i)\right).
  \end{multline*}
  By distributing the product and using the linearity of the trace, we
  get
  \begin{multline}
    \label{eq:expandtrace}
    \tr\left((AB-\sum_{i=1}^k w_i A_i B^i)^T(AB-\sum_{i=1}^k w_i A_i
      B^i)\right) \\
      = \tr\left((AB)^TAB\right) - 2 \sum_{i=1}^k w_i \tr\left((AB)^T
      A_i B^i\right) \\
    + \tr\left(
      (\sum_{i=1}^kw_iA_iB^i)^T(\sum_{i=1}^kw_iA_iB^i)\right),
  \end{multline}
  where we made use of the following equality:
  \begin{align*} \tr\left((AB)^T A_iB^i\right) &=\tr\left(((AB)^T A_iB^i)^T\right) \\
    &=\tr\left((A_iB^i)^TAB\right).
  \end{align*}
  We now work towards rewriting~\eqref{eq:expandtrace} in a more
  manageable form. First, using the fact that $\tr(AB)=\tr(BA)$, we
  see that
  \begin{equation}\label{eq:traceasinnerprod}\tr(A_iB^i)=\langle B^i,
    A_i\rangle.
  \end{equation}

  By combining~\eqref{eq:sumrankone} and~\eqref{eq:traceasinnerprod},
  we have
  \begin{align*}
    \sum_{i=1}^k w_i \tr\left((AB)^T A_iB^i \right) &= \sum_{i=1}^k w_i
    \tr\left(\sum_{j=1}^n(B^j)^T A_j^T A_iB^i\right) \\
    &= \sum_{i=1}^kw_i\sum_{j=1}^n \langle B^i,B^j\rangle\langle
    A_i,A_j \rangle = \<w,r\> \text{.}
  \end{align*}

  Similarly, using~\eqref{eq:traceasinnerprod}, we get after an easy
  computation that
  \begin{multline*}
    \tr\left( (\sum_{i=1}^kw_iA_iB^i)^T(\sum_{i=1}^kw_iA_iB^i)\right)
    \\=\sum_{ij}w_iw_j (\<A_i,A_j\>\<B^i,B^j\> )
    =w^TQ_Jw.
  \end{multline*}

  We now rewrite~\eqref{eq:expandtrace} in a more manageable form:
  \begin{equation}
    \label{eq:expandtracevec}
    \norm{AB-\widetilde{AB}}^2= \norm{AB}^2 -2 \langle w,r\rangle + w^T
    Q_J w.
  \end{equation}

  The optimal weight vector is now obtained by setting the gradient
  of~\eqref{eq:expandtracevec} to zero. Hence we obtain
  \begin{equation*}
    w={Q_J}^{-1}r,
  \end{equation*}
  which proves the first part of the statement.

  For the second part, first notice that if $[\mathbf{1}]$ is the
  vector whose entries are all one, we have the following expression
  for $r$:
  \begin{equation*}
    r=[Q_J \ Y][\mathbf{1}] .
  \end{equation*}

  Hence, at the optimum, the error is
  \begin{align*}
    \norm{AB-\widetilde{AB}}^2 & = \norm{AB}^2-
    [\mathbf{1}]^T
    [Q_J \ Y]^T {Q_J}^{-1}[Q_J \ Y][\mathbf{1}]\\
    & = \norm{AB}^2- [\mathbf{1}]^T \widetilde Q [\mathbf{1}],
  \end{align*}
  where we see that $\widetilde Q$ is the \emph{Nystr\"om approximation
    of} $Q$ as described in Section~\ref{sec:background}.  Using
  Lemma~\ref{lem:normhad} below, we have
  $$
  \norm{AB-\widetilde{AB}}^2=[\mathbf{1}]^T(Q-\widetilde
  Q)[\mathbf{1}],
  $$
  which finishes the proof of the Theorem.

\end{IEEEproof}

The proof of the second part of Theorem \ref{th:maintheo} is based on
the identity proven below:
\begin{Lemma}
  \label{lem:normhad} Let $A$ and $B$ be real matrices of dimensions
  $m \times n$ and $n \times p$, respectively, and let $E$ be the
  $n\times n $ matrix with all entries equal to one.  The following
  identity holds:
  \begin{equation}
    \label{eq:lemmahadamardnorm}\norm{AB}^2=\tr((A^TA)\odot(BB^T)E).
  \end{equation}
\end{Lemma}

\begin{IEEEproof}
  Recall that we can write the product $AB$ as a sum of $n$ rank-one
  terms as follows:
  \begin{equation*}
    AB=\sum_{i=1}^n A_iB^i.
  \end{equation*}
  We thus have, by definition of the Frobenius norm, that
  \begin{eqnarray*}
    \norm{AB}^2&=&\tr((AB)^TAB)\\
    &=&\tr(\sum_i(A_iB^i)^T\sum_j(A_jB^j))\\
    &=&\tr(\sum_{ij}(B^i)^TA_i^TA_jB^j)\\
    &=&\sum_{ij}\tr((B^i)^TA_i^TA_jB^j).
  \end{eqnarray*}

  Using the invariance of the trace with respect to cyclic
  permutations, the last equation yields
  \begin{equation*}
    \norm{AB}^2=\sum_{ij}\<A_i,A_j\>\<B^i,B^j\>,
  \end{equation*}
  and the relation~\eqref{eq:lemmahadamardnorm} is proved.
\end{IEEEproof}

\subsection{Approximating the Optimal Subset Selection Procedure}

Having shown a solution to the optimal reweighting problem according
to Theorem~\ref{th:maintheo}, we now turn our attention to the companion problem
of optimal subset selection.  In order to minimize the approximation error, we have to find the subset $J$ whose associated Schur complement $S_C(Q_J)$ has the lowest possible power along the one-dimensional subspace of $\mathbb{R}^{n-k}$ spanned by the vector $[\mathbf{1}]$.  Determining the eigenvectors and eigenvalues of this Schur complement, and relating them to $A$ and $B$, is not an easy task. Here we present two approximations: one based on a random choice of subsets, and an alternative ``greedy'' approach which yields a worst-case error bound.

\subsubsection{Random Choice of Subset}
\label{sec:randomoracle}

We first discuss a random oracle which outputs a subset $J$ with
probability $p_{Q,k}(J)$ defined below.  Recall our earlier definition of
the matrix $Q=A^TA \odot BB^T$ according to Theorem~\ref{th:maintheo}; this approach
is motivated by the expression of the resultant squared error, conditioned upon having chosen a subset $J \subset \lbrace   1,\ldots,n \rbrace$, as $\norm{AB-\widetilde{AB}}^2 = \tr(S_C(Q_J)E)$. Since $S_C(Q_J)$ is positive definite, we have that $\tr(S_C(Q_J))$  is larger than the largest eigenvalue of $S_C(Q_J)$, and we can bound  this error as follows:
 \begin{equation}
    \label{eq:boundtr}
   \norm{AB-\widetilde{AB}}^2 \leq (n-k) \, \tr(S_C(Q_J)) \text{.}
 \end{equation}
Note that equality is obtained when $S_C(Q_J) \propto E$, and
hence this bound is tight.
We have investigated in~\cite{belabbas:_fast_low_rank_approx_covar_matric}
an algorithm to minimize $\norm{S_C(Q_J)}$, which has been shown to be effective in the context
of low-rank covariance matrix approximation.

Returning to our random oracle, note that both
$A^TA$ and $BB^T$ are positive definite, and thus by the
Schur Theorem~\cite{horn}, $Q$ is also positive definite.  From this, we conclude:
\begin{enumerate}
\item There exists a matrix $X \in \R^{n \times n}$ such that
  $Q=X^TX$;
\item All the principal minors $\det(Q_J)$ of $Q$ are positive.
\end{enumerate}
Consequently, we assume here that an oracle returns a subset $J \subset
\lbrace{1,\ldots,n\rbrace}, |J| = k$ with probability
\begin{equation}\label{eq:defp}
p_{Q,k}(J) := \frac{1}{K}\det(Q_J),
\end{equation}
where $K=\sum_{J,|J|=k}\det(Q_J)$ is a normalizing constant, and the
second fact above ensures that this probability distribution is well
defined. We may then adapt the following result from the proof of Theorem~1 in~\cite{belabbas07:_spect_method_machin_learn}:
\begin{Theorem}\label{th:randomorac}
  Let $Q \in \R^{n \times n}$ be a positive quadratic form with eigenvalues $ \lambda_1 \geq \lambda_2 \geq
  \ldots \geq \lambda_n$.  If $J \subset \lbrace   1,\ldots,n \rbrace,|J|=k$ is chosen with probability
  $p_{Q,k}(J) \propto \det(Q_J)$, then
  \begin{equation}
    \label{eq:avgerrornyst}
  \operatorname{\mathbb{E}} \tr(S_C(Q_J)) \leq (k+1) \sum_{i=k+1}^n \lambda_i \text{.}
  \end{equation}
\end{Theorem}

Combining~\eqref{eq:boundtr} with~\eqref{eq:avgerrornyst} leads, via Jensen's inequality, to an upper bound on the average error of this approach to random subset selection:
  \begin{equation*}
    \operatorname{\mathbb{E}} \norm{AB-\widetilde{AB}}
\leq \sqrt{(n-k)(k+1)} \, \norm{X-X_k} \text{,}
\end{equation*}
where $X$ is defined via the relation $X^TX = A^TA \odot BB^T$, and
$X_k$ denotes the optimal rank-$k$ approximation to $X$ obtained by
truncating its singular value decomposition.

Despite the appearance of the term $\sqrt{n-k}$ in this bound, it serves to relate the resultant approximation quality to the ranks of $A$ and $B$, a feature reinforcing the well-foundedness of the
accompanying algorithm we present below. In particular, if $k>\operatorname{rank}(A)\operatorname{rank}(B)$, then the approximation error is zero as expected. For practical reasons, we may also wish to relate this error to the eigenvalues of $A$ and $B$.  To this end, let $M$
and $N$ be two $n \times n$ matrices, $P=M \odot N$, and let $\sigma_i(M)$ (resp. $\sigma_i(N)$, $\sigma_i(P)$)  be the singular values of $M$  (resp. $N$, $P$) sorted in non-increasing order. We then have the  following majorization relation~\cite{horn2}:
\begin{equation*}
    \sum_{i=1}^m\sigma_i(P) \leq \sum_{i=1}^m
    \sigma_i(M) \, \sigma_i(N), \mbox{ for } m=1,2,\ldots,n.
\end{equation*}

In particular, if $M=A^TA$, $N=BB^T$, and $Q = X^TX=M\odot N$, then
the singular values of $Q$, $M$, and $N$ are the squares of the
singular values of $X,A$ and $B$ respectively:
\begin{equation}
\label{eq:major}
\sum_{i=1}^m\sigma_i^2(X)\leq\sum_{i=1}^m
\sigma_i^2(A)  \, \sigma_i^2(B), \mbox{ for }
m=1,\ldots,n.
\end{equation}
We may then conclude from~\eqref{eq:major} that
\begin{equation*}
\norm{X-X_k}^2 \leq \min\left(\sigma^2_1(A)\norm{B}^2 \,,\, \sigma^2_1(B)\norm{A}^2\right) - \norm{X_k}^2.
\end{equation*}

Although the approach presented above relies on an oracle to sample in proportion to $\det(Q_J)$, we will subsequently outline a realizable algorithm based on these results.

\subsubsection{Deterministic Choice of Subset}

Recall that Theorem~\ref{th:maintheo} indicates we should ensure that
the diagonal terms of $Z$ are kept as small as possible.  Hence, as
a  \emph{deterministic} approximation to the optimal subset
selection procedure, we may take $J$ such that it contains the indices
of the $k$ largest terms $\<A_i,A_i\> \<B^i,B^i\>$.  While yielding
only a worst-case error bound, this approach has the advantage of
being easily implementable (as it does not require sampling according
to $\det(Q_J)$); it also appears to perform well in
practice~\cite{belabbas:_fast_low_rank_approx_covar_matric}.  This
greedy algorithm proceeds as follows:
\begin{algorithm}
 \caption{Greedy Approximate Matrix Multiplication}
 \label{alg:pseudocodemult}
 Given matrices $A \in \R^{m \times n}$ and $B \in \R^{n \times p}$
 and a positive integer $k\leq n$:
  \begin{enumerate}
  \item Set $T:=\left\lbrace \<A_{i},A_{i}\>\<B^{i},B^{i}\>
    \right\rbrace$, $i=1,\ldots,n$, and take $J:=\left\lbrace
      i_1,\ldots i_k\right\rbrace$ to be the indices of the $k$
    largest elements of $T$.
  \item Set $Q \in \R^{k \times k}$ as
    $Q_{ij}:=\<A_{i},A_{j}\>\<B^{i},B^{j}\>$, for $i,j \in J$.
  \item Set $r \in \R^k$ as $r_i:=\sum_{j=1}^n
    \<A_{i},A_{j}\>\<B^{i},B^{j}\>$, for $i \in J$.
  \item Set $w:=Q^{-1}r$ and $A_{J}:=\{A_i\}, B_{J}:=\{B^i\}$ for $i \in J$.
  \item Return $\widetilde{AB}:=A_{J}\operatorname{diag}(w)B_{J}$ as an
  approximation to $AB$.
  \end{enumerate}
\end{algorithm}

Since the error term is the sum of all the terms in the Schur complement, we
can look to bound its largest element. To this end, we have the
following result:
\begin{Lemma}
  \label{prop:greedy}
  The largest entry in $S_C(Q_J)$ is smaller than the largest diagonal
  element of $Z$ in~\eqref{eq:partQ}.
\end{Lemma}
This lemma confirms that a good error-minimization strategy is to make sure that
the diagonal terms of $Z$ are as small as possible, or equivalently
to take $J$ such that it contains the indices of the $k$ largest
$\<A_i,A_i\> \<B^i,B^i\>$ as per Algorithm~\ref{alg:pseudocodemult}.

\begin{figure}[!t]
  \centering
  \includegraphics[width=\columnwidth]{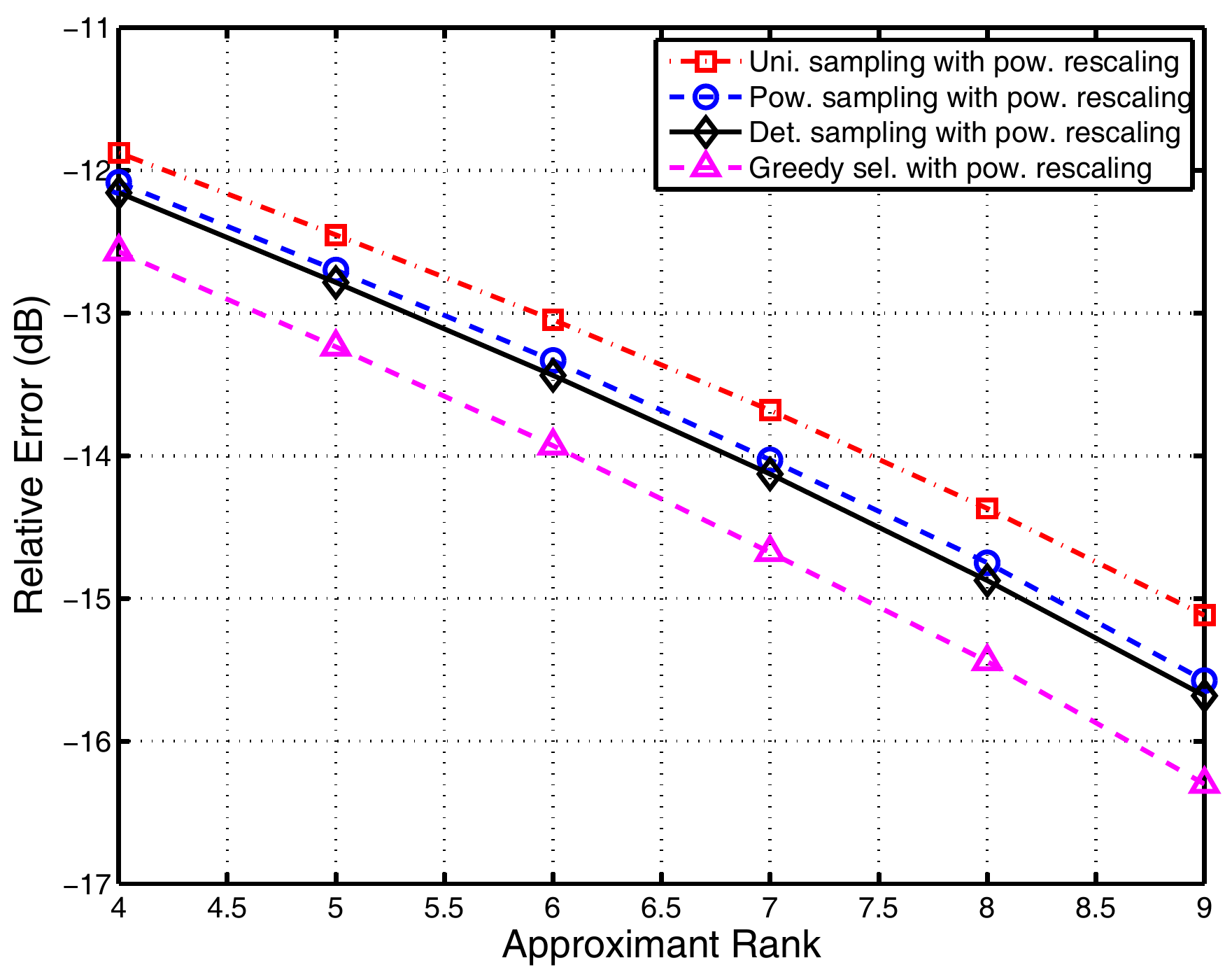}
  \caption{\label{fig:fig1}Matrix product approximation error using the power rescaling of~\eqref{eq:powerscale} applied to each of the four subset selection algorithms described in Sec.~\ref{sec:results}}
\end{figure}

The proof of Lemma~\ref{prop:greedy} is based on the following
set of simple results:
\begin{Lemma}
  \label{lem:lardiag}
  If $Q$ is a positive definite matrix, then $\max_{ij}Q_{ij}$ is positive and on the diagonal of
  $Q$.
\end{Lemma}
\begin{IEEEproof}
  Since $Q$ is positive definite, we know there exists a matrix $X \in
  \R^{n\times n}$ such that
  \begin{equation*}
    Q_{ij}=\<X_i,X_j\>.
  \end{equation*}
  By the Cauchy-Schwartz inequality, we have
  \begin{equation*}
    \<X_i.X_j\>^2 \leq
    \<X_i,X_i\>\<X_j,X_j\>,
  \end{equation*}
  from which we deduce that one of the following inequalities has to
  be satisfied:
  \begin{equation}
    \label{eq:altercauchy}
    \<X_i,X_j\> \leq \<X_i,X_i\> \mbox{ or }
    \<X_i,X_j\> \leq \<X_j,X_j\>.
  \end{equation}

  Now if we suppose that $\max_{ij}Q_{ij}$ is not a diagonal element,
  the relations of~\eqref{eq:altercauchy} yield a contradiction---and
  hence the largest entry of $Q$ is on its main diagonal.
\end{IEEEproof}

The entries of $S_C(Q_J)$, the Schur complement of $Q_J$ in $Q$, can
be characterized explicitly according to the following formula:
\begin{Lemma}[Crabtree-Haynsworth~\cite{crabtree69:_ident_schur_compl_matrix}]
  \label{lem:crabtree}
  Let $Q_J=Q_{1,\ldots,k;1\ldots,k}$ be a nonsingular leading
  principal submatrix of $Q$ obtained by keeping the rows and columns
  with indices $1,\ldots,k$. Then $S_C(Q_J)$, the Schur complement of
  $Q_J$ in $Q$, is given element-wise by
  \begin{equation}
    (S_C(Q_J))_{ij}=\frac{\det(Q_{1,\ldots,k,i;1,\ldots,k,j})}{\det(Q_J)}.
  \end{equation}
\end{Lemma}
Furthermore, it is possible to bound the diagonal entries of $S_C(Q_J)$ as
follows:
\begin{Lemma}[Fischer's Lemma~\cite{horn}]
\label{lem:Fisch}
  If $Q_J$ is a positive definite matrix, then
  \begin{equation*}
    \det(Q_{J\cup \{i\}})
    < \det(Q_J)Q_{ii}.
  \end{equation*}
\end{Lemma}

We are now ready to give the proof of Lemma~\ref{prop:greedy}:
\begin{IEEEproof}[Proof of Lemma~\ref{prop:greedy}]
  The preceding two lemmas tell us that the diagonal entries of
  $S_C(Q_J)$ are bounded by $\max_{i \notin J} Q_{ii}$ (i.e., the
  largest diagonal element of $Z$, according to the partition
  of~\eqref{eq:partQ}).  And using Lemma~\ref{lem:lardiag}, we know
  that every entry of $S_C(Q_J)$ is bounded by these diagonal entries.
\end{IEEEproof}

Lemma~\ref{prop:greedy} can be further refined to give
a worst-case error bound for deterministic matrix product
approximation, conditioned on a choice of subset $J$ and the
corresponding optimal reweighting procedure.
Appealing to the inequality of arithmetic and geometric means to further bound the elements of $S_C(Q_J)$, the results of Theorem~\ref{th:maintheo} and Lemmas~\ref{lem:lardiag}--\ref{lem:Fisch} yield:
\begin{equation*}
  \norm{AB-\widetilde{AB}} \leq \sqrt{(n-k) \sum_{i \notin
    J}(\norm{A_i}^2\norm{B^i}^2)} \text{.}
\end{equation*}

\begin{figure}[!t]
  \centering
  \includegraphics[width=\columnwidth]{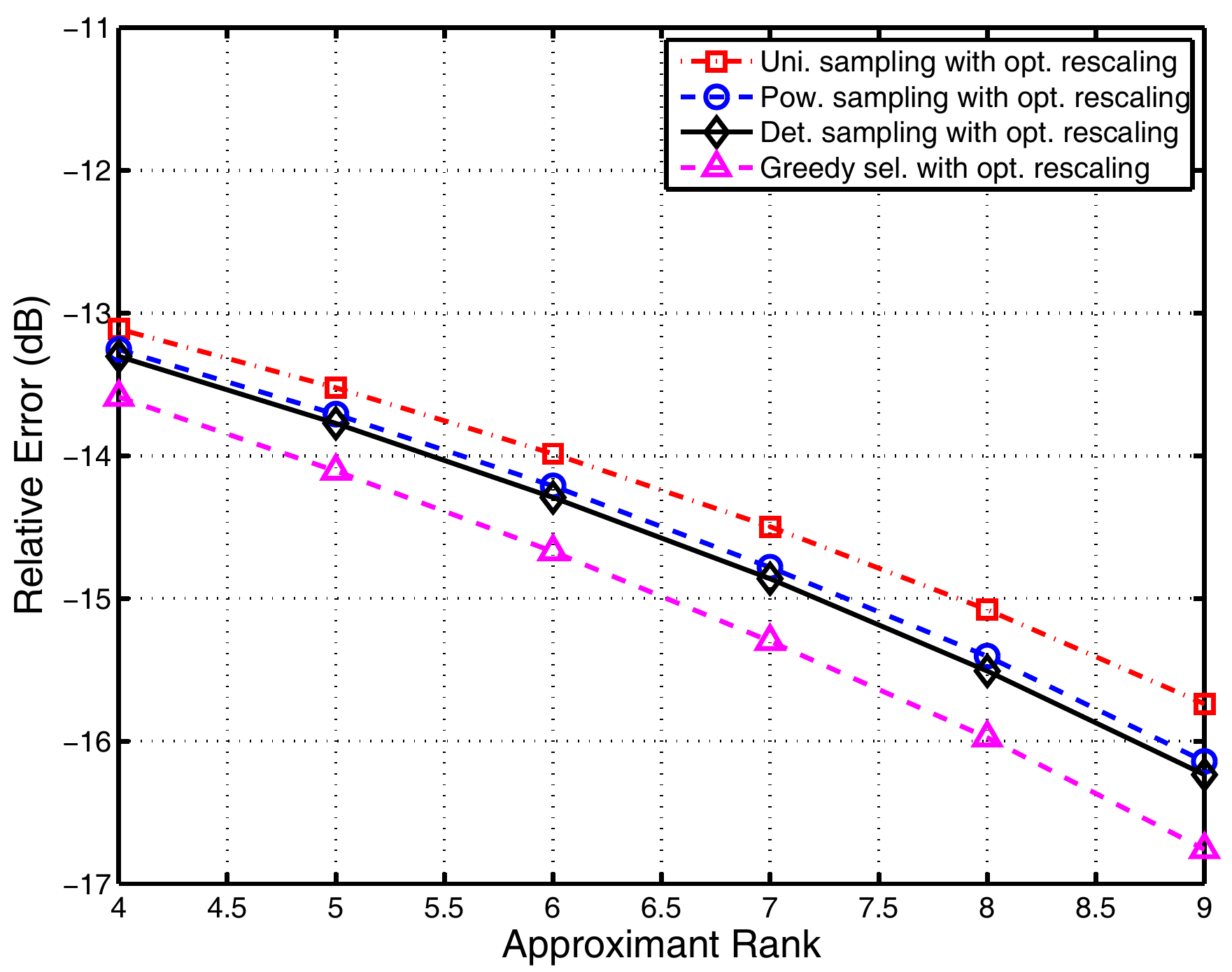}
  \caption{\label{fig:fig2} Matrix product approximation error using the optimal rescaling of Theorem~\ref{th:maintheo} applied to the subset selection algorithms described in Sec.~\ref{sec:results}}
\end{figure}

\section{Simulation Studies and Complexity}
\label{sec:perf_complexity}

\subsection{Experimental Results}
\label{sec:results}

We now present preliminary experimental results and discuss the
computational complexity of the algorithms under consideration. Three sets of experiments were performed, in which we compared the performance of four subset selection methods: a baseline uniform sampling on $k$-subsets; sampling according the row/column powers~\cite{drineas06fast}; sampling in proportion to the $k$-principal minors of $Q$ according to~\eqref{eq:defp}; and selecting greedily according to Step~1 of Algorithm~\ref{alg:pseudocodemult}.  We also compared the choice of reweighting following subset selection, in one case applying the optimal reweighting of Theorem~\ref{th:maintheo} and in the other simply reweighting according to the row/column powers (see~\cite{frieze98fast, drineas06fast}):
\begin{equation}
\label{eq:powerscale}\widetilde{AB} = \sum_{i \in J} \frac{1}{\sqrt{|J|\norm{A_i}^2\norm{B^i}^2}} A_iB^i.
\end{equation}

To test these various experimental conditions, we drew 200 random matrices $A \in \R^{60 \times 15}$ and $B \in \R^{15 \times 90}$ in total, each having independent unit Normal entries.  We then averaged the error of the randomized algorithm over 20 trials per matrix product, and report relative error in~dB as $20 \log_{10} \left(\norm{AB-\widetilde{AB}} / \left(\norm{A}\norm{B}\right)\right)$ for each test condition.

In the first set of experiments, shown in Figure~\ref{fig:fig1}, we compare the four different algorithms for subset selection described above, applied in conjunction with a reweighting according to row/column powers.  The highest-error method in this case corresponds to choosing the subset $J$ uniformly at random, and thus should be understood as a baseline measure of performance as a function of approximant rank $k$.  It can also be seen that sampling $J$ according to the relative powers of the row/columns of $A$ and $B$, and sampling via a Metropolis-Hastings algorithm (with independent proposal distributions taken in proportion to row/column powers), yield similar results, with both improving upon the baseline performance.  The best results in this case are obtained by the greedy subset selection method indicated by Step 1 of Algorithm~\ref{alg:pseudocodemult}.

In a second set of experiments, we followed the same procedure as above to compare subset selection procedures, but applied the optimal reweighting of Theorem~\ref{th:maintheo} rather than a rescaling according to row/column powers. Performance in this case is (as expected) seen to be better overall,  but with the ordering of the methods unchanged.  As our final experiment, we compare the method of Algorithm~\ref{alg:pseudocodemult} (greedy subset selection followed by optimal rescaling) to two non-adaptive methods: choosing row/columns of $A$ and $B$ uniformly at random and rescaling according to $n/k$, and the simple Johnson-Lindenstrauss random projection approach outlined in Section~\ref{sec:intro}.
These non-adaptive methods can be seen to yield significantly worse performance than Algorithm~\ref{alg:pseudocodemult}, suggesting its potential as a practical method of selecting sparse representations of linear operators that yield low approximation errors for the resultant matrix products.

We conclude with a brief discussion of the algorithmic complexity of Algorithm~\ref{alg:pseudocodemult}.  First, assume without loss of generality that $m \geq n,p$, and recall that
straightforward matrix multiplication requires $\Or(m^3)$ operations,
though the best algorithm known so far (the Coppersmith-Winograd
algorithm~\cite{cw-mmap-90}) can perform this operation in
$\Or(m^{2.38})$.  Evaluating $T$ in
Algorithm~\ref{alg:pseudocodemult} requires the computation of $2n$
inner products of $m$-or $p$-dimensional vectors, and hence requires
$\Or(2mn)$ operations.  Extracting the $k$ largest elements of a set
of size $m$, as is necessary to construct $J$, can be done efficiently
using a variation on the Quicksort algorithm (see~\cite{rivest}) in
$\Or(m \log k)$. The matrix $Q$ is symmetric and its diagonal is a
restriction of $T$.  Hence it requires the computation of an
additional $2\times k(k-1)/2$ inner products, and thus $\Or(mk(k-1))$
operations.  Evaluating $r$ requires $\Or(2m(n-k))$ operations, taking
into account the fact that $k$ terms of the sum also appear in $Q$.
Finally, evaluating $w$ can be done using Gaussian elimination in
$\Or(k^3)$ operations. Hence the overall complexity is given by
$\Or(m(k(k-1)+2(2n-k)+\log k)+k^3)=\Or(m(n+k^2)+k^3)$.

\begin{figure}[!t]
  \centering
  \includegraphics[width=\columnwidth]{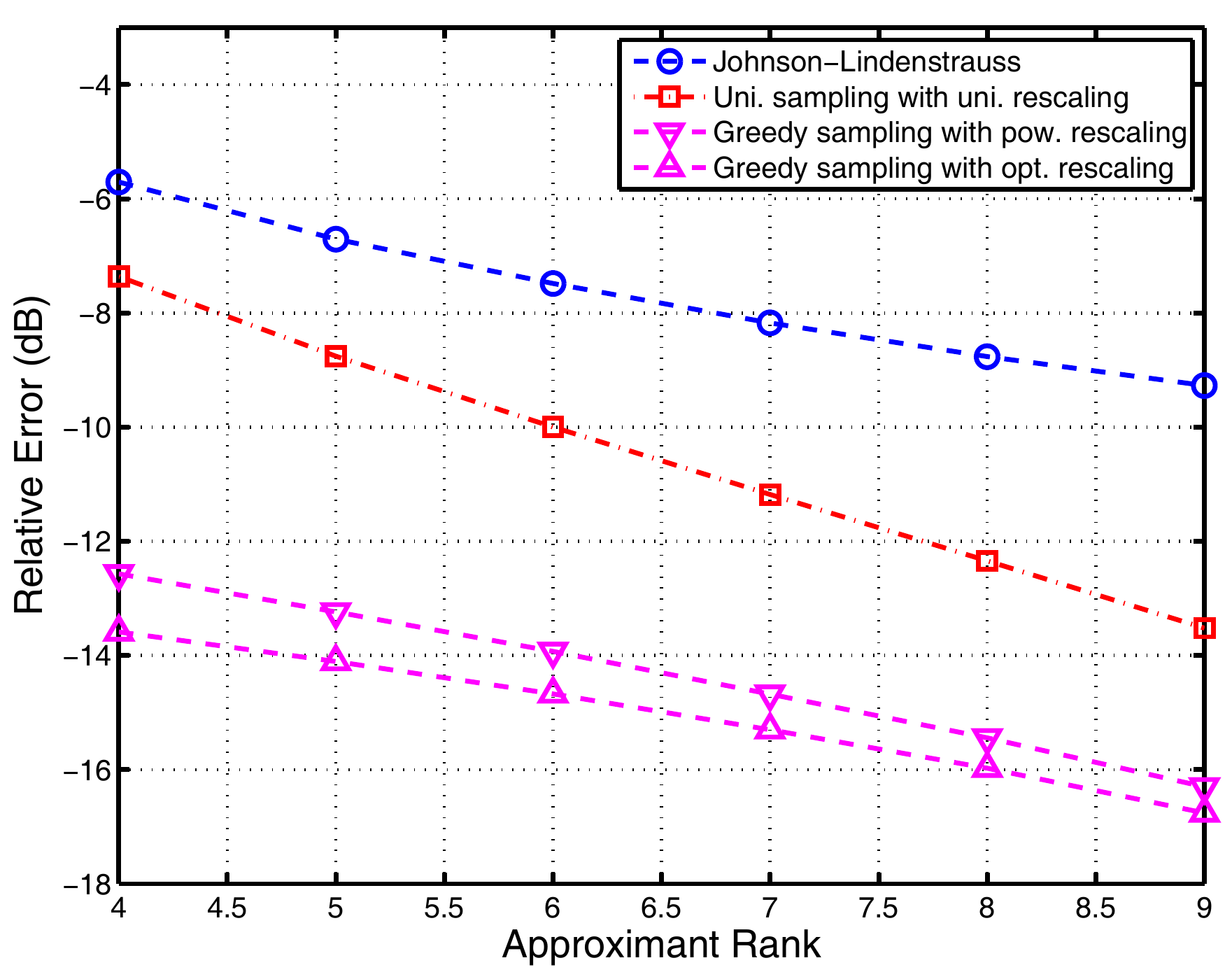}
  \caption{\label{fig:fig3} Matrix product approximation error using non-adaptive random projections (Johnson-Lindenstrauss), non-adaptive subset selection (uniform), and adaptive subset selection (Algorithm~\ref{alg:pseudocodemult})}
\end{figure}

\bibliographystyle{IEEEbib}

\end{document}